\documentclass[aps,prl,reprint,superscriptaddress]{revtex4-2}
\usepackage[T1]{fontenc}
\usepackage[latin9]{inputenc}
\usepackage{color}
\usepackage{amsmath}
\usepackage{graphicx}
\usepackage[english]{babel}
\usepackage{hyperref}
\usepackage{float}

\newcommand{\mr}{\mathrm}
\newcommand{\pd}{{\phantom{\dagger}}}

\usepackage{blkarray}
\usepackage{bigstrut}

\begin{document}

\title{Resolving space-time structures of quantum impurities with a numerically exact few-body algorithm}

\author{Yuriel N\'u\~nez Fern\'andez}
\affiliation{Univ. Grenoble Alpes, CNRS, Grenoble INP, Institut N\'eel, 38000 Grenoble, France}
\author{Maxime Debertolis}
\affiliation{Institute of Physics, University of Bonn, Nussallee 12, 53115 Bonn, Germany}
\author{Serge Florens}
\affiliation{Univ. Grenoble Alpes, CNRS, Grenoble INP, Institut N\'eel, 38000 Grenoble, France}

\begin{abstract}
We introduce a numerically exact real-time evolution scheme for quantum impurities in a macroscopically large bath. 
The algorithm is \emph{few-body revealing}, namely it identifies the electronic orbitals that can be made inactive 
(in a trivial product state) by a time-dependent orbital rotation.
Following a quench, we show that both the number of active orbitals and their associated matrix product state bond 
dimensions saturate to small values, leading to an algorithm dramatically more accurate and faster than the state of the art. 
We are thus able to follow the dynamics for thousands of fermions, up to the long-time stationary regime, and to study 
subtle aspects of quantum relaxation in the spatio-temporal domain, such as the emergence of entanglement structures 
in the Kondo screening cloud.
\end{abstract}

\maketitle

The many-body dynamics of generic quantum systems currently challenges numerical simulations, as
exemplified by the unbounded entanglement growth in time of matrix product states (MPS). This
situation also plagues quantum impurity problems, ``simpler'' systems still comprised of a
macroscopic number of particles (bosons or fermions), where the interaction is restricted to a
subset of degrees of freedom~\cite{Costi_RMP}. The interest in impurity models, originating in Kondo
physics, has not waned over the years, since they are the cornerstone of the Dynamical Mean-Field
Theory (DMFT) for correlated electrons~\cite{Georges_RMP}. Open quantum systems, in which qubits
exchange energy and information with an environment, constitute also a very important class of
impurity models~\cite{LeHur_Impurity}. However, in most cases, simulating the real-time
non-perturbative dynamics of impurity models comes with tremendous numerical cost: typical runs
taking over a week only achieve a limited $1\%$ accuracy on observables for quenches, and relatively
short reachable time scales~\cite{Costi_Dynamics,Costi_Quench}. Progress is often made by losing the
information over the spatial dynamics of the
environment~\cite{Anders_TDNRG,Cohen_Memory,Millis_TimeEvolution,Werner_Continuous,Goldstein_Lindblad,
Yuriel_TCI,Abanin_InfluenceFunctional,Abanin_MPS,Chan_InfluenceFunctional}.

We pose in this Letter the idea of \emph{few-body revealing} for quantum impurities models, based on
a new finding that their quench dynamics involves only a small number of entangled environmental
degrees of freedom, called {\it active orbitals}, that continuously adapt during the time evolution.
The rest of the macroscopic environment remains at each time in a trivial, yet evolving, product
state (Slater determinant for fermions), making the simulation of systems with thousands of sites
now reachable. Entanglement growth is prevented by collecting at each time step the orbital rotation
into an {\it evolving unitary frame} that is never applied to the state (this would make the problem
many-body again), but only to observables. Our main advance is to propose a {\it numerically exact
algorithm} (with controllable errors below $10^{-6}$) that takes advantage of this few-body
structure, leading to a drastic reduction of the computing time (down to only minutes for full
relaxation of systems with hundreds of sites), while giving access to real-space information on the
environment. This algorithm allows us to study the buildup of the dynamical Kondo cloud after a
quench, uncovering new aspects of decoherence in a fermionic environment.

\begin{figure}
\begin{centering}
\includegraphics[width=1.\columnwidth]{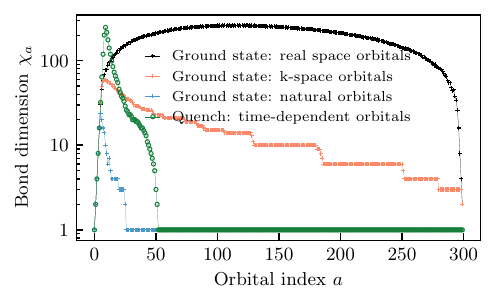}
\par\end{centering}
\caption{Demonstrating \emph{few-body revealing} from the MPS bond dimension $\chi_a$ as a function
of orbital index $a$, for the IRLM with $L=300$ sites at $U/D=-0.5$ and $V/D=0.1$. For ground states
(crosses), natural orbitals show a drastic compression w.r.t. real-space and k-space orbitals,
because all natural orbitals with $\chi_a=1$ are contained in a Slater state. For the dynamics
following a quench (circles), an optimal orbital rotation leaves the state vector $|\psi(t)\rangle$
few-body at the long time $Dt=100$, while $\chi_a$ would exceed here $10^4$ for real-space and
k-space, making the computation unfeasible.}
\label{fig1}
\end{figure}
\paragraph{Model.}
We consider the Interacting Resonant Level model
(IRLM)~\cite{Vigman_IRLM} given by the real-space Hamiltonian
in terms of local fermion operators $\hat{c}_{i}^{\dagger}$ 
and charges $\hat{n}_i=\hat{c}^{\dagger}_i\hat{c}^\pd_i$:
\begin{eqnarray}
\nonumber
\hat{H} &=& U\Big(\hat{n}_{0}-\frac{1}{2}\Big)\Big(\hat{n}_{1}-\frac{1}{2}\Big)
+\epsilon_{0}(t)(\hat{n}_{0}-\hat{n}_1)\\
&&+ V (\hat{c}^{\dagger}_0 \hat{c}^\pd_1+\hat{c}^{\dagger}_1 \hat{c}^\pd_0)
+ \frac{D}{2}\sum_{i=1}^{L-2}(\hat{c}_{i+1}^{\dagger}\hat{c}^\pd_{i}+\mr{h.c.}),
\label{eq:H}
\end{eqnarray}
with interaction $U$, hybridization $V$, and hoppings $D/2$.
The time-dependent impurity onsite energy will be quenched from $\epsilon_{0}(t)=-\infty$ at $t<0$
(so that sites $0$ and $1$ are initially decoupled from the bath), to $\epsilon_{0}(t)=0$ at $t>0$. 
The IRLM is a very good testbed for quantum impurities since it exhibits Kondo correlations at $U<0$ 
in the charge sector.

A first insight onto the problem is given by the matrix product state (MPS) bond dimension distribution $\chi_a$ 
for the ground state of the IRLM (at fixed $\epsilon_0=0)$, see Fig.~\ref{fig1}.
We observe a huge compression when using the natural orbital basis, namely
when solving Hamiltonian~(\ref{eq:H}) in the set of fermionic operators 
$\sum_i U_{ai}\hat{c}^\pd_i$ that diagonalize 
the correlation matrix $\rho_{ij}=\langle\hat{c}_{i}^{\dagger}\hat{c}^\pd_{j}\rangle$.
Not only the bond dimension is much smaller for natural orbitals than for real-space and k-space
orbitals, but also the number of active orbitals is less than 40 (this is nearly independent of the
system size), meaning that almost all the orbitals are in a perfect product state (Slater
determinant). This is in agreement with previous studies based on iterative
diagonalization~\cite{He_Lu_NO,Maxime_FewBody,Maxime_Cloud} and non-Gaussian
states~\cite{Florens_Snyman_Bosonic,Bravyi_Gosset,Cirac_Demler,Snyman_Florens_Fermionic,Boutin_Bauer}.
One important finding of this Letter is that a time-dependent orbital rotation maintains this
few-body picture at any time after a local quench, as can be seen from the curve of Fig.~\ref{fig1}
in circle symbols. We now present the algorithm allowing us to obtain controlled calculations for
the quench at long times, taking advantage of the few-body picture. This algorithm can be
naturally extended beyond the IRLM (for instance with fermions carrying spin and orbital indices),
since the few-body character generally results from the locality of the impurity degrees of freedom
immersed in a macroscopic environment. Our code is publicly available on the "FastQuantum" 
repository~\cite{FastQuantum}, and further developments will be added to library.

\paragraph{General frame evolution.}
We propose that the physical state at a given time $t$ is given by 
$\exp[-i \hat{F}(t)]|\psi(t)\rangle$, where
$\hat{F}(t)=\sum_{a,b=0}^{L-1}F_{ab}(t)\hat{c}_{a}^{\dagger}\hat{c}^\pd_{b}$
is a quadratic form in terms of fermion operators, parametrized by an hermitian one-body matrix 
$F_{ab}(t)$ that we call {\it frame}. Previous studies have shown disentangling properties of 
orbital rotations~\cite{Haverkort_Solver,Schollwock_DMFT,Legeza_PRL,Legeza_Rotation,Santoro_Thermofield,Moreno_Rotation,Fishman_Disentangle}, 
but were limited by a fixed time-independent frame, or did not take full advantage of the simplifications 
brought by impurity models.
Picking an optimal frame $F_{ab}(t)$ will keep a very low entanglement in the many-body 
wavefunction $|\psi(t)\rangle$ for the whole time evolution.
This is possible because for the quench dynamics of impurity models, only a small 
number of active orbitals 
$N_\mr{active}\ll L$ is involved in $|\psi(t)\rangle$, as hinted in Fig.~\ref{fig1}.
From Thouless theorem~\cite{Thouless_Theorem}, the change of frame transforms each orbital
to a linear combination as
\begin{equation}
\exp[i \hat{F}(t)]\hat{c}^\pd_{a}\exp[-i \hat{F}(t)]=\sum_{b}\big[e^{-i F(t)}\big]_{ab}\hat{c}^\pd_{b}\text{.}
\label{eq:orb_rot}
\end{equation}
Importantly, we restrict the rotation $F_{ab}$ only to the bath orbitals (namely to sites 
$i>1$ in our model), in order to leave the interaction term of the impurity unaffected.

\paragraph{Trotterized evolution.}
We notice that the application of a general orbital rotation $\exp[-i\hat{F}(t)]$ to 
a given MPS $|\psi(t)\rangle$ is as expensive as a full time evolution, and should be 
avoided at all costs.
The first trick is to separate the time evolution between two parts: a macroscopic part 
stemming only from the {\it bath} variables (namely sites $i=2,\ldots,L-1$), that is never 
applied to the state $|\psi(t)\rangle$, and a microscopic part that is efficiently applied 
to $|\psi(t)\rangle$ using shallow quantum gates. Evolution is performed by splitting
the bath Hamiltonian
$\hat{H}_\mr{bath}\equiv
(D/2)\sum_{i=2}^{L-2}\big(\hat{c}_{i}^{\dagger}\hat{c}^\pd_{i+1}+\mr{h.c.}\big)$
from the impurity part $\hat{H}_\mr{imp}\equiv\hat{H}-\hat{H}_\mr{bath}$. While the operator
$\hat{H}_\mr{bath}$ accounts for most of the energy of the system for large $L$, it is a quadratic
Hamiltonian whose associated time evolution corresponds to an orbital rotation that can be
incorporated into the frame $F_{ab}(t)$. To isolate the macroscopic bath operator from the impurity
one, we perform a Trotter time evolution on a small interval $dt$ using the third order
Baker--Campbell--Hausdorff formula:
\begin{equation}
e^{-i \hat{H}dt} = e^{-i\hat{H}_\mr{bath}dt -i\hat{H}_\mr{imp}dt}
\approx e^{-i \hat{H}_\mr{bath}dt}
e^{-i \hat{H}_\mr{imp}^{(2)}dt}\text{.}
\label{eq:int_pic}
\end{equation}
The first term is incorporated to a new frame $F(t+dt)$, and only
the second term drives the time evolution with a dressed Hamiltonian
$\hat{H}_\mr{imp}^{(2)}=\hat{H}_\mr{imp}-idt[\hat{H}_\mr{imp},\hat{H}_\mr{bath}]/2$.
$\hat{H}_\mr{imp}^{(2)}$ is expressed in 
the current frame $F(t)$, which rotates only bath orbitals.
The interaction term $U\hat{n}_0\hat{n}_1$ and hybridization 
$V (\hat{c}^{\dagger}_0 \hat{c}^\pd_1+\hat{c}^{\dagger}_1 \hat{c}^\pd_0)$ are 
both invariant under bath rotations,
and only the coupling term $(D/2) \hat{c}^{\dagger}_1 \hat{c}^\pd_2$ (and its order $dt$
correction) is affected.
After frame rotation according to Eq.~(\ref{eq:orb_rot}), $\hat{H}_\mr{imp}^{(2)}$ takes the form:
\begin{equation}
\label{eq:few_body_hamil}
\hat{H}_\mr{imp}^{(2)} = U \hat{n}_0\hat{n}_1 
+ V (\hat{c}^{\dagger}_0 \hat{c}^\pd_1 + \hat{c}^{\dagger}_1 \hat{c}^\pd_0) 
+ \sum_{a,b=0}^{L-1}K_{ab} \hat{c}^{\dagger}_a \hat{c}^\pd_b.
\end{equation}
For the IRLM, the matrix $K_{ab}$ couples only site $i=1$ to a linear combination 
of bath fermions, and is therefore rank 1 (in general its rank is the number 
of electronic channels coupling to the impurity).

\paragraph{Few-body evolution.}
We re-express Hamiltonian~(\ref{eq:few_body_hamil})
in terms of the orbital $\hat{f}^\pd_2$ coupling to $c^\dag_1$ through $K_{ab}$:
\begin{equation}
\hat{H}_\mr{imp}^{(2)} = U \hat{n}_0\hat{n}_1 
+ V (\hat{c}^{\dagger}_0 \hat{c}^\pd_1 + \hat{c}^{\dagger}_1 \hat{c}^\pd_0) 
+ K (\hat{c}^{\dagger}_1 \hat{f}^\pd_2 + \hat{f}^{\dagger}_2 \hat{c}^\pd_1).
\label{eq:three_body_hamil}
\end{equation}
Therefore, the time evolution will be trivial (few-body) if we can efficiently rotate 
the MPS state $|\psi(t)\rangle$ into this appropriate basis.
This can be achieved by an intra-bath rotation $Q_\mr{few}$ that is
associated to the first right singular vector of $K_{ab}$ 
(see Ref.~\cite{Fishman2015} and Sec.~I of Ref.~\cite{supp}).

In practice, we also need to perform this rotation so that its action on the
bath keeps the state few-body. The trick is to
classify at a given time bath orbitals into three groups according to their occupation number
$n_{a}=\langle\hat{c}_{a}^{\dagger}\hat{c}^\pd_{a}\rangle$:
\emph{empty} ($0\!\leq\! n_{a}\!\leq\!\epsilon$),
\emph{full }($1-\epsilon \!\leq\! n_{a} \!\leq\! 1$), and \emph{active} ($\epsilon<n_{a}<1-\epsilon$),
with $\epsilon$ a small cutoff (typically $\epsilon=10^{-10}$ or less).
At order $\epsilon$, we can thus consider that empty/full orbitals are strictly inactive,
namely $n_a=0$ or $n_a=1$, so that any rotation of the empty or full orbitals (that does not
mix them) now leaves the corresponding Slater determinant invariant.
By doing the rotation $Q_\mr{empty}$ (resp. $Q_\mr{full}$) that extracts the first right singular vector 
of matrix $K_{ab}$ in the empty (resp. full) sector, we obtain that all empty (full) orbitals are isolated 
except the first one.
If the state at a given time $t$ had $N_\mr{active}$ active orbitals, the evolved state 
at time $t+dt$ will have $N_\mr{active}+2$ active ones, and all inactive orthogonal 
bath orbitals will still remain in a (rotated) Slater state.

\paragraph{Orbitals optimization.}
After evolving the state $|\psi(t)\rangle$, its optimal frame has changed 
and must be adapted to reduce the entanglement within the new set of $N_\mr{active}+2$ 
active orbitals.
A set of local rotations $Q_\mr{opt}$ can diagonalize the correlation matrix
$\rho_{ab}=\langle\psi|\hat{c}_{a}^{\dagger}\hat{c}^\pd_{b}|\psi\rangle$ among
the $N_\mr{active}+2$ active orbitals only, since the remaining $L-(N_\mr{active}+2)$
orbitals are either empty or full.
The diagonalization is done as proposed in~\cite{Fishman2015} for Slater states. We use
nearest neighbor Givens rotations~\cite{Anderson2000GivensRot} to extract the extremum eigenpair of 
the correlation matrix (see Sec.~I of Ref.~\cite{supp}), 
and repeating this process up to visiting all eigenpairs.
After this step, the number of active orbitals most often goes back to the previous value
$N_\mr{active}$ only, and its total increase over the whole temporal simulation will assess
how few-body the state has stayed. These local rotations are finally mapped to local
quantum gates that are efficiently applied to the state 
$|\psi(t)\rangle$, bringing it in the new frame.

\paragraph{The few-body revealing algorithm.}
To summarize, our algorithm operates by simple iterative updates of the
frame matrix $F_{ab}(t)$, and by a few-body evolution of the state $|\psi(t)\rangle$ 
for each time step $dt$, as follows:
\begin{enumerate}
\item[(I)] \emph{Trotterized evolution}: The frame is updated as 
$\exp(-iF)\leftarrow \exp(-i H_\mr{hop} dt)\exp(-iF)$ from the Trotter
decoupling, where 
$[H_\mr{hop}]_{ij}=(D/2)\delta_{i,j\pm1}$ is the hopping matrix appearing
in $\hat{H}_{\mr{bath}}$.
\item[(II)] \emph{Few-body evolution}: 
The frame is updated as 
$\exp(-i F)\leftarrow\exp(-i F) Q_\mr{empty}Q_\mr{full}Q_{\text{few}}$,
and we map $Q_\mr{few}$ to a shallow quantum circuit that we apply to the state $|\psi(t)\rangle$.
The state is then time-evolved by block decimation~\cite{White_TDDMRG,ITensor2022} upon
$\hat{H}_\mr{imp}^{(2)}$ that acts only on three orbitals, and $N_\mr{active}+2$ orbitals are now active.
\item[(III)] \emph{Orbital optimization}: The frame is updated as 
$\exp(-iF)\leftarrow\exp(-i F)Q_{\text{opt}}$, and $Q_\mr{opt}$ is mapped
as a shallow quantum circuit to the state $|\psi(t)\rangle$. The number of active orbitals 
most often drops back to $N_\mr{active}$, and the entanglement is reduced. 
\end{enumerate}
The Trotter error in step~(I) is $\mathcal{O}(dt^{3})$, due to Eq. (\ref{eq:int_pic}).
Since the full time evolution consists in $(t/dt)$ steps, the expected Trotter error on observables 
at a finite time $t$ scales like $\mathcal{O}(dt{}^{2})$ when $dt\to0$. 
Importantly, the Trotter error in our algorithm is purely local, and does not scale with system size, 
contrarily to the Trotter dynamics based on a real-space chain, see Sec.~III.B
in~\cite{supp}. Other approaches, {\it e.g.} based on the MPS compression of influence 
functionals~\cite{Abanin_MPS} also show increasing numerical costs when trying to diminish the Trotter error.
Step~(II) introduces an error $\mathcal{O}(\epsilon)$ from the threshold $\epsilon$ used to freeze 
the nearly empty and full orbitals. We additionally compress the MPS with the same tolerance $\epsilon$ 
after applying the gates in steps (II-III). 
All errors are controlled and can be systematically suppressed without dramatically increasing
costs, so that our algorithm is numerically exact.
To compute the expectation value of a given operator at any time, we can trivially 
rotate each fermion appearing in it to the current frame according to Eq.~(\ref{eq:orb_rot}), 
since the frame is given by a quadratic Hamiltonian. Note that step (I) alone would be the traditional 
interaction picture of the bath, and would lead to entanglement growth. Only the addition of steps (II-III) 
provides an algorithm that reveals the few-body and low-entanglement character of the underlying state.
\begin{figure}[ht]
\begin{centering}
\includegraphics[width=1.\columnwidth]{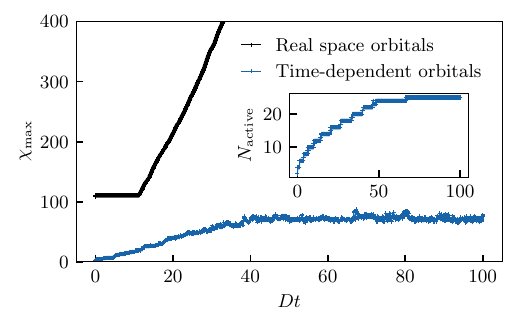}
\par\end{centering}
\caption{Time-evolution of \emph{few-body revealing}, seen from the maximum bond dimension
$\chi_\mr{max}$, for the quench dynamics of IRLM at $U/D=0.2$, $V/D=0.1$, $L=200$ sites, and Trotter
step $dt=0.1$. Our algorithm using time-dependent orbital rotation shows bounded entanglement at
long times, while the numerical costs diverge for simulations with real-space orbitals.
\emph{Inset}: the total number of active orbitals (with $\chi_a> 1$) is also seen to saturate over
time, and dynamics remains few-body.}
\label{fig2}
\end{figure}
\paragraph{Results.}
We first begin by illustrating the remarkable disentangling feature of our \emph{few-body revealing}
algorithm for a quench of the IRLM. Figure~\ref{fig2} shows the time evolution 
of the maximum bond dimension $\chi_\mr{max}=\mr{Max}[\chi_a]$, comparing two Trotter-based calculations 
using either the time-dependent orbital basis or the real-space basis. We observe that $\chi_\mr{max}$ 
saturates to small values in the former, while it grows unbounded in the latter. We find similar entanglement 
growth using TDVP dynamics in the star-basis~\cite{FastQuantum}. In addition, the number of active orbitals 
$N_\mr{active}$ in the new algorithm remains bounded at long times, making the scheme controlled and efficient.

\begin{figure}
\begin{centering}
\includegraphics[width=1.\columnwidth]{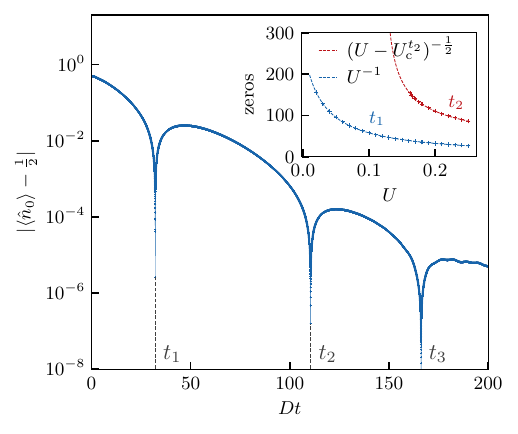}
\par\end{centering}
\caption{Impurity charge relaxation toward the equilibrium value $n_0(t\to\infty)=1/2$ for the IRLM at $U/D=0.2$, $V/D=0.1$, $L=200$ sites, 
and Trotter step $dt=0.01$. The dynamics for $U>0$ is coherent, with a leading exponential envelope, and a certain number of zeros indicated as $t_1, t_2, t_3$ here. \emph{Inset}: the first zero $t_1$ diverges as $1/U$ for $U\to0$ (in agreement with analytical results extracted from Ref.~\cite{Kennes_Zeroes}), while the second zero $t_2$ is found to diverge at finite $U_\mr{c}^{t_2}/D\simeq0.12$ with a different 
power law $(U-U_\mr{c}^{t_2})^{-\frac{1}{2}}$.
}
\label{fig3}
\end{figure}
A crucial test of the accuracy of our new algorithm is given in Fig.~\ref{fig3}. Previous time-dependent NRG and DMRG studies~\cite{Costi_Dynamics} for the same quench protocol have shown that errors could not be decreased below $10^{-2}$, 
preventing a detailed study of charge relaxation at long times. 
Interestingly, it was predicted~\cite{Kennes_Zeroes} that the damping of the charge imbalance at the impurity site, $n_0(t)-1/2$, would crossover from a fully coherent regime (with an infinite number of zeros) at $U>U_\mr{c}$, to a partially coherent regime (with a finite number of zeros) for $0<U<U_\mr{c}$. Such coherent regime, here displaying three zeros at times $t_1$, $t_2$, $t_3$, is shown in the main panel of Fig.~\ref{fig3}. We extracted these zeros for many $U$ values (see inset), and we observe that the first zero $t_1$ disappears at $U\to0$ as expected from the existence of the incoherent regime at $U\leq0$~\cite{Leggett_Review}, and diverges as $1/U$, in agreement with the analytics of Ref.~\cite{Kennes_Zeroes}. Interestingly, we find that the subsequent zero $t_2$ diverges at a finite $U_\mr{c}^{t_2}$, a new observation implying that relaxation occurs via a unique zero $t_1$ 
for $0<U<U_\mr{c}^{t_2}$. This is compatible with (yet clearly beyond) the predictions of Ref.~\cite{Kennes_Zeroes}.
In addition, we are able to identify a different power law behavior
$t_2\propto(U-U_\mr{c}^{t_2})^{-\frac{1}{2}}$ governing the disappearance of the second zero. Our
study surmounts a numerical challenge, because it requires controlling all errors down to $10^{-6}$,
as detailed in Sec.~III.A of Ref.~\cite{supp}.

\begin{figure}[ht]
\begin{centering}
\includegraphics[width=1.0\columnwidth]{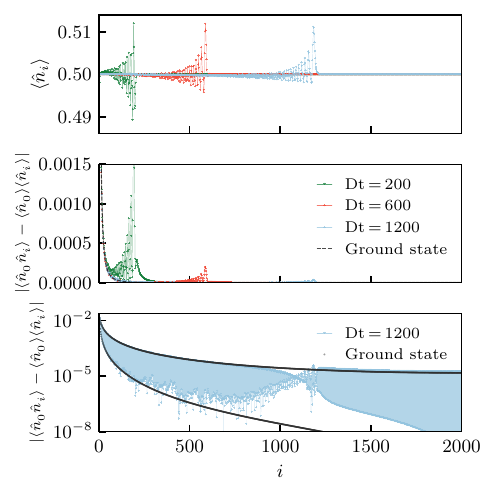}
\par\end{centering}
\caption{\emph{Top}: Local charge $\langle\hat{n}_i\rangle$ on site $i$ for a long chain of $L=2000$ sites, with $U/D=-0.2$, $V/D=0.1$, and Trotter step $dt=0.1$, showing the ballistic motion of the charge wavepacket 
at three different times.
\emph{Middle}: charge correlation function $|\langle \hat{n}_0 \hat{n}_i\rangle 
- \langle \hat{n}_0 \rangle \langle\hat{n}_i\rangle|$ for the same parameters as the top panel, only on even sites.
At distances $i\ll Dt$, a charge Kondo cloud binds progressively to the impurity. 
At distances $i\simeq Dt$, a ballistic peak of correlations between the local charge $\hat{n}_i$ in the 
wavepacket and the impurity charge $\hat{n}_0$ is seen, that decays in amplitude as $t\to\infty$, due to 
loss of entanglement outside the Kondo cloud. 
\emph{Bottom:} same plot as the middle panel, but for both even/odd sites, now on a logarithmic scale and only 
for the time $Dt=1000$.
The comparison to the ground state solution (lines) shows proper relaxation to the Kondo cloud for $i\ll Dt$,
and also the formation of a dynamical cloud for $i\gtrsim Dt$, despite the fact that the excess charge is vanishing 
after the wavefront (top panel).}
\label{fig4}
\end{figure}

We finally investigate the spatial structures in the environment at long times after the quench for
the IRLM at negative $U/D=-0.2$ (Kondo regime, corresponding to a large relaxation time $Dt\simeq
100$), for a very large system with $L=2000$ sites. Such dynamics can be easily computed in the
non-interacting case~\cite{Eisler_Dynamics,supp}, but is very challenging in presence of
interactions. Focusing on the local charge $\langle\hat{n}_i\rangle$ (top panel of
Fig.~\ref{fig4}), we see that the extra density moves away ballistically from the impurity located
at site $i=0$, without any damping or dispersion. Spatial Kondo correlations, given by the
observable $|\langle \hat{n}_0 \hat{n}_i\rangle - \langle \hat{n}_0 \rangle
\langle\hat{n}_i\rangle|$, have been the topic of intense studies for the ground state of impurity
models~\cite{Affleck_Review}. While Numerical Renormalization Group (NRG) simulations are feasible
in and out of equilibrium~\cite{Borda_Cloud,Anders_Cloud}, they require the expensive construction
of a double bath for each computed distance in the chain, while our simulation is done with only one
bath, and the computation of any observable is single shot. Errors are also maintained below
$10^{-4}$ on all sites in our calculation. For the quench, we find several remarkable observations
in these spatial correlations. First, at short distances $0<i\ll Dt$ compared to the wavepacket, a
Kondo cloud bound to the impurity progressively forms, see middle/bottom panels of Fig.~\ref{fig4}
(for linear/logarithmic views). This constitutes a non-trivial check that the relaxation to
equilibrium is correctly established in our numerics. Second, we observe a peak of correlations on
sites that take part in the wavefront, namely for $i\lesssim Dt$ (middle panel). The emitted charge
wavepacket, although it is spread over hundred of sites, maintains a finite entanglement to the
impurity. Relaxation to equilibrium is nevertheless reached at $t\to\infty$, since these
correlations decay at long times, while the excess charge remains constant in amplitude. Third, we
reveal new details on the Kondo-like correlations that emerge in front of the wavepacket for sites
at distances $i\gtrsim $ (bottom panel). While ``supra-luminous'' correlations are
expected~\cite{Anders_Cloud}, they remarkably reach very long distances beyond the wavefront, in a
region of space where charge neutrality is exactly balanced, and quantitatively differ from the
Kondo cloud correlations that are bound to the impurity. A full movie of the dynamics is available
in the results section of the "FastQuantum" repository~\cite{FastQuantum}.

\paragraph{Outlook.} Using optimal time-dependent orbital rotations, we showed that \emph{few-body
revealing} makes the quench dynamics of quantum impurities almost trivial (namely few-body and low
entangled). This finding gives rise to a numerically exact algorithm that is very simple to
implement, faster and more accurate that previous state of the art, and able to cope with
macroscopically large systems and long times. The idea of time-dependent orbital rotation can 
be tailored to other frameworks, for instance we have checked that Lanczos-based dynamics highly benefits 
from orbital rotation, and it would be interesting to investigate if influence functionals~\cite{Abanin_MPS} 
could obtain compression gains as well. Remarkably, orbital rotation at the super-operator level
can be used to compress the full evolution operator of impurity models in a matrix product 
representation~\cite{Debertolis_Operator}.

Our advance opens obviously many promising perspectives for the dynamical simulations 
of interacting quantum systems. For spinful Anderson
quantum impurities, the spatio-temporal dynamics of spin-charge separation could be explored,
possibly at finite temperature with a thermofield approach~\cite{Santoro_Thermofield}, or under
finite bias transport condition~\cite{StarBasisTransport}. Generalizing orbital rotations to
superconducting quantum dots~\cite{Paaske_SCdot} should also open the possibility to study the
dynamics of interacting Andreev bound states. Interaction effects in electronic 
waveguides~\cite{Bauerle_Waveguide}, often neglected~\cite{Kloss_Pulse,Gouraud_NESS}, could 
also be investigated.
Another promising route concerns real-time quantum impurity solvers~\cite{Schollwock_DMFT} for the
DMFT equations~\cite{Georges_RMP}, especially in the multi-orbital case~\cite{Evertz_ForkTensor},
and in the context of out-of-equilibrium physics~\cite{Werner_TDDMFT}, a fascinating and challenging
topic~\cite{Schiro_Thermal}. Finally, extensions of time-dependent orbital rotation to bosons and
two-level systems could help the simulation of driven superconducting qubits~\cite{Blais_Transmon}
and quantum processing platforms~\cite{Vidal_TPU}.

\emph{Acknowledgements}.
YNF and SF were supported by the PEPR integrated project EPiQ ANR-22-PETQ-0007 part of Plan France 2030.
MD was supported by the Deutsche Forschungsgemeinschaft through the cluster of excellence ML4Q (EXC 2004, 
project-id 390534769) and by the Deutsche Forschungsgemeinschaft through CRC 1639 NuMeriQS (project-id 511713970).
We also thank T.Costi, V. Meden, and I. Snyman for stimulating discussions.

\bibliographystyle{apsrev4-2}
%

\setcounter{figure}{0}
\setcounter{table}{0}
\setcounter{equation}{0}

\global\long\def\theequation{S\arabic{equation}}
\global\long\def\thefigure{S\arabic{figure}}
\renewcommand{\thetable}{S\arabic{table}}
\renewcommand{\arraystretch}{0.6}

\normalsize

\begin{center}
{\bf \large Supplemental material for ``Resolving space-time structures of quantum impurities 
with a numerically exact few-body algorithm}
\end{center}

\section{Extracting single orbital using quantum gates}

In this section, we explain how we can construct a convenient set of quantum
gates that acts on the state vector in order to move a targeted orbital 
onto the first site, adapting ideas from Ref.~[33] of the main text. Let us start 
by presenting the method of Givens rotations (see Ref.~[35] of the main text 
for more details) in the simplest case of a two component complex vector 
$\boldsymbol{v} = (v_1,v_2)^T$ such that $|v_1|^2+|v_2|^2=1$.
We define the $2\times2$ complex and unitary "Givens" matrix as:
\begin{equation}
G =\begin{pmatrix}c & s\\
-\bar{s} & c
\end{pmatrix}\text{,}
\label{2by2Givens}
\end{equation}
with $c=|v_1|$ and $s=v_1 \bar{v_2}/|v_1|$, 
denoting here $\bar{s}$ the complex conjugate. We can clearly eliminate 
the second component $v_2$ of $\boldsymbol{v}$ by acting 
with the Givens matrix:
\begin{equation}
G\cdot\begin{pmatrix}v_1\\
v_2   
\end{pmatrix}=\begin{pmatrix}r\\
0
\end{pmatrix}
\end{equation}
with $r=f/|f|$ a pure phase.

The first step of our algorithm is to transform algebraically a fixed targeted orbital, 
associated to a normalized eigenvector $\boldsymbol{v}=(v_{1},v_{2},\ldots,v_{L})^T$, into the
simple elementary vector $\boldsymbol{e}_{1}=(1,0,0,\ldots,0)^T$. 
Up to a global phase, this is performed by using a succession of simple Givens rotations:
\begin{equation}
G_{1}G_{2}\ldots G_{L-1}\boldsymbol{v}
\equiv G\cdot\boldsymbol{v}=\boldsymbol{e}_{1}\text{,}
\label{GivensProduct}
\end{equation}
using $L\times L$ Givens matrices defined as:
\begin{equation}
G_i = \;\;
\setlength{\bigstrutjot}{1ex}
\begin{blockarray}{ccccccccc}
  & & & & & & & & \\
    \begin{block}{(>{\medspace}cccccccc<{\medspace})r}
1 & 0 & \ldots & 0 & 0 & \ldots & 0 & 0 & \bigstrut[t] \\
0 & 1 & \ldots & 0 & 0 & \ldots & 0 & 0 & \\
0 & 0 & \ldots & c & s & \ldots & 0 & 0 &  \; \leftarrow i\phantom{\hspace{0.15cm}+1} \\
0 & 0 & \ldots & -\bar{s} & c & \ldots & 0 & 0 & \hspace{0.2cm} \leftarrow i+1 \\
0 & 0 & \ldots & 0 & 0 & \ldots & 1 & 0 & \\
0 & 0 & \ldots & 0 & 0 & \ldots & 0 & 1 & \bigstrut[b]\\
\end{block}
\end{blockarray}
\label{FullGivens}
\end{equation}
so that $G_{i}$ acts only on the rows $i$ and $i+1$ of $\boldsymbol{v}$, 
eliminating the component $i+1$. The rationale behind Eq.~(\ref{GivensProduct})
is therefore to eliminate all components of $\boldsymbol{v}$ from $v_L$ up to $v_2$.

The second step of our algorithm is to apply the Givens rotation to achieve the extraction
of a targeted eigenvector $\boldsymbol{v}$ of the correlation matrix
$\rho_{ij}=\langle\psi(t) | \hat{c}_i^\dagger \hat{c}^\pd_j |\psi(t)\rangle$
computed at a given time $t$ from the many-body MPS wavefunction $|\psi(t)\rangle$. 
Let us note the corresponding eigenvalue equation as $\rho \boldsymbol{v} = n \boldsymbol{v}$,
with $\boldsymbol{v}$ the natural orbital associated to the occupation $n$. We can focus
on the active space of size $M\ll L$ where $\epsilon \leq n \leq 1-\epsilon$, with a small 
cutoff $\epsilon$ introduced to freeze all nearly inactive orbitals (see main text). 
Using the fact that $G$ is an unitary matrix, 
and $G \rho \boldsymbol{v} = n G \boldsymbol{v}$, we get
$G \rho G^\dagger (1,0,0,\ldots,0)^T = n (1,0,0,\ldots,0)^T$.
Thus we obtain the general form:
\begin{equation}
G.\rho.G^\dagger =
  \left[ \begin{array}{c|ccc}
    n & 0  & \ldots & 0 \\
    \hline
    0 & & & \\
    \vdots & \multicolumn{3}{c}
       {\raisebox{\dimexpr\normalbaselineskip-.7\ht\strutbox-.5\height}[0pt][0pt]
        {\scalebox{3}{$\rho'$}}}  \\
    0 & & &
  \end{array} \right]
\end{equation}
where the matrix $\rho'$ has one column and one row less than matrix $\rho$.
Repeating this process for all $M$ active orbitals amounts to fully diagonalize
$\rho$ in the active space. By construction, the remaining $L-M$ orbitals are
fully empty or full in the associated wavefunction $|\psi(t)\rangle$, and do 
not require extraction.

The last step of our algorithm is to translate the Givens rotation~(\ref{GivensProduct})
associated to the extraction of a targeted orbital into one layer of elementary quantum gates
acting on the many-body wavefunction $|\psi(t)\rangle$. This makes the change of frame computationally
manageable. For this purpose we associate to the Givens rotation $G_i$ defined in 
Eq.~(\ref{FullGivens}) a quantum gate $\hat{G}_i$ that acts only on the Fock basis 
\{$\{|00\rangle,|10\rangle,|01\rangle,|11\rangle\}$ for the two sites $i$ and $i+1$ involved 
in the rotation of the full wavefunction $|\psi(t)\rangle$. 
It is straighforward to check that a two-sites orbital rotation such as Eq.~(\ref{2by2Givens}) 
amounts at the wavefunction level to the following four-by-four gate acting on the local
Fock states:
\begin{equation}
G_{i}\longrightarrow\hat{G}_{i}\equiv\begin{pmatrix}1 & 0 & 0 & 0\\
0 & c & -\bar{s} & 0\\
0 & s & c & 0\\
0 & 0 & 0 & 1
\end{pmatrix}\text{,}
\label{GivensMapping}
\end{equation}
where $\hat{G}_i$ acts as identity on the rest of the full Hilbert space that
does not contain orbitals $c_i$ and $c_{i+1}$. 
The rotation of the many-body MPS wavefunction corresponding to the extraction 
of a single eigenvalue of $\rho$ is implemented through a sequence of quantum 
gates obtained from the sequence Eq.~(\ref{GivensProduct}) and the mapping~(\ref{GivensMapping}).
Applying these rotations one after the other to all $M$ active orbitals results in a many-body 
wavefunction where the active and inactive sectors are exactly decoupled.

\section{Exact dynamics in the non-interacting case}
In order to test carefully our general MPS algorithm, we derive an exact formal solution 
for the resonant level model, corresponding to the non-interacting case of the problem 
studied in the main text.
We start with the free real-space Hamiltonian:
\begin{eqnarray}
\label{eq:H0}
\hat{H_0} &=& V(c^\dag_0 c^\pd_1 + c^\dag_1 c^\pd_0)
+\frac{D}{2}\sum_{j=1}^{L-1}\big(\hat{c}_{j}^{\dagger}\hat{c}^\pd_{j+1}+\mr{h.c.}\big)\\
\nonumber
&=& V(\hat{c}^\dag_0\hat{c}^\pd_1 + \hat{c}^\dag_1\hat{c}^\pd_0)
+\sum_{k=2}^{L-1} \epsilon_k \tilde{c}^\dagger_k \tilde{c}^\pd_k
+V' \sum_{k=2}^{L-1}\big( \hat{c}_1^{\dagger} \tilde{c}^\pd_k +\mr{h.c.}\big)
\end{eqnarray}
expressing the fermionic chain operators in Fourier space as
$\hat{c}^\pd_j = (L-2)^{-1/2}\sum_{k=2}^{L-1}\tilde{c}^\pd_k e^{i 2\pi (j-2) (k-2)/(L-2)}$ for
sites $j=2,\ldots,L-1$. The coupling to the bath now reads
$V'=(L-2)^{-1/2} D/2$ and the one-particle energy of the chain is given by
$\epsilon_k = -D\cos[2\pi (k-2)/(L-2)]$ for $k=2,\ldots,L-1$.

The initial state $|\psi(0)\rangle$ at time $t=0$ considered in the main text
corresponds to having site $j=0$ occupied, site $j=1$ empty, and
the rest of the chain decoupled from the impurity, which therefore 
reads as a product state involving the $\tilde{c}^\pd_k$ orbitals with negative energy
(at half-filling):
\begin{equation}
\label{eq:initial}
|\psi(0)\rangle = \hat{c}_0^\dagger \prod_{\epsilon_k<0} \tilde{c}_k^\dagger |0\rangle.
\end{equation}
We relabel all the fermionic fields as $\hat{f}^\pd_a$, with
$\hat{f}^\pd_0=\hat{c}^\pd_0$,
$\hat{f}^\pd_1=\hat{c}^\pd_1$,
$\hat{f}^\pd_a=\tilde{c}^\pd_a$ (for $a=2,\ldots,L-1)$ for compactness.

We now define the one-body (correlation) matrix
at a given time $t$ as $\rho_{ab}(t)=\langle\psi(t) |
\hat{f}_a^\dagger \hat{f}^\pd_b |\psi(t)\rangle$,
which reads in Heisenberg picture $\rho_{ab}(t)=\langle\psi(0) |
\hat{f}_a^\dagger(t) \hat{f}^\pd_b(t) |\psi(0)\rangle$,
with $\hat{f}_a^\pd(t) = e^{i\hat{H}_0 t} \hat{f}^\pd_a e^{-i\hat{H}_0 t}$,
where $\hbar=1$ has been set.
We now express the free Hamiltonian~(\ref{eq:H0}) in its eigenbasis,
$\hat{H}_0 = \sum_{p=0}^{L-1} \Omega_p \hat{\gamma}_p^\dagger \hat{\gamma}^\pd_p$,
with the unitary transformation to operators $\hat{\gamma}^\pd_p 
= \sum_{a=0}^{L-1} U_{pa}\hat{f}^\pd_a$, that can be obtained numerically
for a cost $O(L^3)$ by diagonalizing the one-particle Hamiltonian $h_{ab}$
with matrix elements
$h_{aa} = \epsilon_a \theta(a-1)$,
$h_{a<b} = V\delta_{a0}\delta_{b1}+(D/2)\delta_{a1}\theta(b-1)$,
and $h_{a>b} = h_{a<b}$. 
The time-evolution is now explicitly expressed as:
\begin{eqnarray}
\hat{f}^\pd_a(t) &=& \sum_{p=0}^{L-1} U_{pa}^* 
e^{i\hat{H}_0 t} \hat{\gamma}^\pd_p e^{-i\hat{H}_0 t}
= \sum_{p=0}^{L-1} U_{pa}^* e^{-i\Omega_p t} \hat{\gamma}^\pd_p\,\,\,\,\\
\nonumber
&=& \sum_{a'=0}^{L-1} \sum_{p=0}^{L-1} U_{pa}^* e^{-i\Omega_p t} 
U_{pa'}\hat{f}^\pd_{a'} \equiv \sum_{a'=0}^{L-1} M_{aa'}(t) 
\hat{f}^\pd_{a'}.
\end{eqnarray}
We thus obtain the correlation matrix at time $t$ as:
\begin{equation}
\rho_{ab}(t) = \sum_{a',b'=0}^{L-1} M_{aa'}^*(t) M_{bb'}(t)
\langle\psi(0) |
\hat{f}_{a'}^\dagger \hat{f}^\pd_{b'} |\psi(0)\rangle.
\label{rhoFinal}
\end{equation}
For the initial state
$|\psi(0)\rangle = \hat{f}_0^\dagger \prod_{\epsilon_a<0} \hat{f}_a^\dagger |0\rangle$,
the average $\langle\psi(0) | \hat{f}_{a'}^\dagger \hat{f}^\pd_{b'} |\psi(0)\rangle
= n_{a'} \delta_{a'b'}$, with $n_{a'}=0$ or $1$ the initial occupation of orbital $a'$. 
We finally obtain the real space correlation matrix $\langle c^\dagger_i c_j\rangle$ by 
back Fourier transform of Eq.~(\ref{rhoFinal}).

\section{Benchmarking errors}
\subsection{Comparison to the exact free solution}
\begin{figure}[ht]
\begin{centering}
\includegraphics[width=1.0\columnwidth]{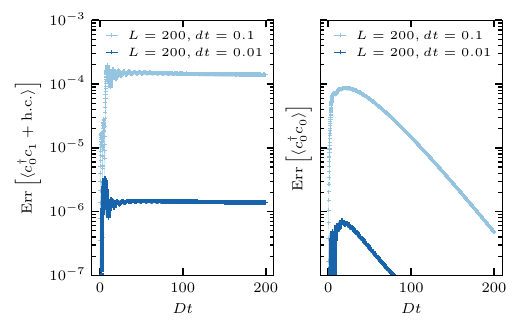}
\par\end{centering}
\caption{Numerical errors for the impurity hybridization (left panel) and 
the local charge (right panel) as a function of time, for the quench of 
the non-interacting resonant level model
for $V/D=0.1$ and $L=200$ sites.
The values from the exact solution~(\ref{rhoFinal}) were subtracted to the
MPS results. The drop from $10^{-4}$ down to $10^{-6}$ by diminishing the
time step from $dt=0.1$ to $dt=0.01$ confirms the expected $(dt)^2$ scaling of
the Trotter error.}
\label{figSup1}
\end{figure}

We now compare results obtained from our numerical algorithm to
the exact solution at zero interaction. In Figure~\ref{figSup1}, we show the impurity
hybridization $\langle\hat{c}^\dagger_0\hat{c}^\pd_1+\hat{c}^\dagger_1\hat{c}^\pd_0\rangle
=\rho_{01}+\rho_{10}$ (left panel) and 
charge $\langle\hat{c}^\dagger_0\hat{c}^\pd_0\rangle=\rho_{00}$ (right panel)
as a function of time $t$ for two different Trotter discretizations $dt=0.1$ and $dt=0.01$.
The exact result from Eq.~(\ref{rhoFinal}) is subtracted to reveal the
convergence of the numerics.
Even for the relatively large time step $dt=0.1$, errors are
already as small as $10^{-4}$. Errors go down to $10^{-6}$ for
$dt=0.01$, confirming the general scaling $(dt)^2$ of the Trotter
discretization.

\begin{figure}[H]
\includegraphics[width=1.0\columnwidth]{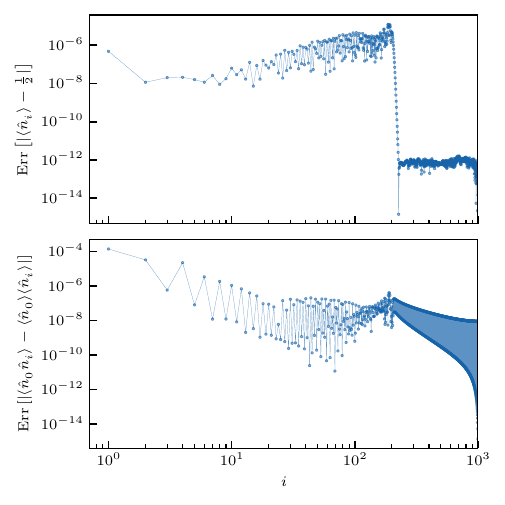}
\caption{Errors on the local charge (top panel) and on the Kondo charge correlator 
(bottom panel) for a very long chain of $L=1000$ sites at fixed time $Dt=200$, 
with the same parameters as Fig.~\ref{figSup1}, showing also systematic convergence in the
spatial domain.}
\label{figSup2}
\end{figure}

We can also establish the convergence of the spatial structures. In the top panel 
of Fig.~\ref{figSup2}, we plot the error on the local charge $\langle \hat{n}_i\rangle$
along the chain for a very long system with $L=1000$ sites, computed in the
non-interacting case $U=0$ for $V/D=0.1$, at fixed time $Dt=200$ with a Trotter 
time step $dt=0.1$.
Again we have subtracted the exact solution Eq.~(\ref{rhoFinal}) to the numerical MPS
simulations.
The bottom panel shows the error on the Kondo cloud charge correlator
$|\langle \hat{n}_0 \hat{n}_i\rangle - \langle \hat{n}_0 \rangle \langle\hat{n}_i\rangle|$.
For the exact free solution, we can use Wick's theorem, and we find
$|\langle \hat{n}_0 \hat{n}_i\rangle - \langle \hat{n}_0 \rangle \langle\hat{n}_i\rangle|
= |\langle \hat{c}_0^\dagger \hat{c}^\pd_i\rangle|^2 = |\rho_{0i}|^2$. For the
MPS simulation, the cloud is computed without use of Wick's theorem, as discussed 
in the main text.
Both plots in Fig.~\ref{figSup2} show errors at the maximum level of $10^{-4}$, 
demonstrating the complete control of the algorithm over the full many-body wavefunction 
(and not just on impurity observables).

\subsection{Comparison to real space DMRG}
\begin{figure}[H]
\includegraphics[width=1.0\columnwidth]{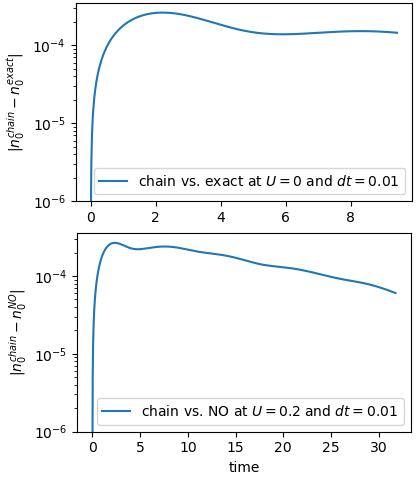}
\caption{Top panel: Comparison of the chain dynamics for the impurity charge $n_0(t)$ with respect
to the exact solution at $U=0$ (here for a chain of $L=100$ sites and small Trotter step $dt=0.01$).
Larger errors than in the orbital rotation scheme are observed.
Bottom panel: Comparison of the chain dynamics to the orbital rotation algorithm for finite
interaction $U=0.2$. The same magnitude of errors as the free case indicate that errors are
not strongly dependent on interactions.}
\label{figSup3}
\end{figure}

For comparison purposes, it is instructive the assess the accuracy of 
the time-evolution obtained with the standard real space approach,
based on Trotter evolution of the fermionic chain. We emphasize that 
it is numerically very costly to reduce the Trotter step, compared to our 
natural orbital algorithm, and for this reason, we will only study 
the dynamics at short times.
In the top panel of Fig.~\ref{figSup3}, we compare the chain dynamics at $U=0$ 
to the exact result Eq.~(\ref{rhoFinal}), for the smallest Trotter time
step $dt=0.01$ and $L=100$ sites. We see that the error stabilizes at order 
$10^{-4}$, which is much larger than the typical error of $10^{-6}$ obtained
using the orbital rotation method for the same $dt$ value, see 
Fig.~\ref{figSup1}. The main reason is that the orbital rotation method
is based on a local Trotter decoupling, while it occurs in the bulk of the
chain for the real space method, thereby leading to a factor $L$ in the global
error in the latter. Nevertheless, the chain dynamics is well under control at short times
and for chains with moderate lengths, and we do find that its accuracy improves 
when the Trotter steps are reduced, as expected. However, maintaining this 
level of error at longer times in the chain dynamics would involve diverging 
numerical costs due to the increase of bond dimension. 

For finite interactions, we do not have access to an exact solution, but we can benchmark
the chain dynamics and the orbital rotation algorithm with respect to each other,
see the bottom panel of Fig.~\ref{figSup3}, performed for $U=0.2$ (other
values of $U$ provide similar results) and $L=100$ sites. Here we see again the 
same $10^{-4}$ mismatch, and we can easily understand that the chain dynamics is prone 
to the same errors as found in the free case. While we cannot confirm the $10^{-6}$
accuracy of the orbital rotation scheme based on this comparison (but at most a $10^{-4}$
mismatch), we can assess independently the errors of our algorithm by both a systematic 
decrease of the Trotter steps, and by examining the typical noise level of the observables
(see Fig. 3 of the main text), noting that the origin of all errors are well identified and
controlled in our scheme.

\end{document}